\documentclass[twocolumn,
                prl,showpacs,%
                superscriptaddress]{revtex4}

\usepackage{graphicx}
       \newcommand{\iv}{\emph{IV} }%
       \newcommand{\tn}{\left \lbrace \tau_n \right \rbrace}%
\usepackage{amsmath}
\usepackage{amssymb}
\usepackage[colorlinks=true]{hyperref}
\usepackage[slantedGreek]{mathptmx}
\begin{document}

\title{
       Distribution of conduction channels in nanoscale contacts:
       evolution towards the diffusive limit
        }
       \author{\href{mailto:juanjo.riquelme@uam.es}{J. J. Riquelme}}
       \affiliation{Laboratorio de Bajas Temperaturas, Dept. F\' isica de
       la Materia Condensada C-III, Universidad Aut\'onoma de Madrid,
       E-28049 Madrid, Spain.}
       \author{L. de la Vega}
       \affiliation{Dept. F\' isica Te\'orica de la Materia Condensada
       C-V, Universidad Aut\'onoma de Madrid, E-28049 Madrid, Spain.}

       \author{A. Levy Yeyati}
       \affiliation{Dept. F\' isica Te\'orica de la Materia Condensada
       C-V, Universidad Aut\'onoma de Madrid, E-28049 Madrid,
       Spain.}\affiliation{Instituto Universitario de Ciencia de
       Materiales ``Nicol\' as Cabrera'', Universidad Aut\'onoma de
       Madrid, E-28049 Madrid, Spain.}

       \author{N. Agra\"{i}t}
       \affiliation{Laboratorio de Bajas Temperaturas, Dept. F\' isica de
       la Materia Condensada C-III, Universidad Aut\'onoma de Madrid,
       E-28049 Madrid, Spain.} \affiliation{Instituto Universitario de
       Ciencia de Materiales ``Nicol\' as Cabrera'', Universidad
       Aut\'onoma de Madrid, E-28049 Madrid, Spain.}

       \author{A. Martin-Rodero}
       \affiliation{Dept. F\' isica Te\'orica de la Materia Condensada
       C-V, Universidad Aut\'onoma de Madrid, E-28049 Madrid,
       Spain.}\affiliation{Instituto Universitario de Ciencia de
       Materiales ``Nicol\' as Cabrera'', Universidad Aut\'onoma de
       Madrid, E-28049 Madrid, Spain.}

       \author{G. Rubio-Bollinger}
       \affiliation{Laboratorio de Bajas Temperaturas, Dept. F\' isica de
       la Materia Condensada C-III, Universidad Aut\'onoma de Madrid,
       E-28049 Madrid, Spain.} \affiliation{Instituto Universitario de
       Ciencia de Materiales ``Nicol\' as Cabrera'', Universidad
       Aut\'onoma de Madrid, E-28049 Madrid, Spain.}

\date{\today}

\begin{abstract}
    We present an experimental determination of the conduction channel
    distribution in lead nanoscale contacts with total conductances ranging
    from 1 to 15 $G_0$, where $G_0 = 2e^2/h$.
    It is found that even for contacts having a cross section much smaller
    than the mean free path the distribution tends to be remarkably close to
    the universal diffusive limit. With the help of theoretical calculations
    we show that this behavior can be associated with the specific band
    structure of lead which produces a significant contribution of partially
    open channels even in the absence of atomic disorder. Published
    in Europhysics Letters,
    \href{http://www.edpsciences.org/articles/epl/abs/2005/11/epl8748/epl8748.html}
    {http://www.edpsciences.org/articles/epl/abs/2005/11/epl8748/epl8748.html}.
\end{abstract}

\pacs{73.23.-b, 73.63.Rt, 73.40.Jn}

\maketitle


    Electron transport properties in a quantum coherent conductor
    are fully characterized by the set of transmission coefficients
    $\tn$ corresponding to the conductance eigenchannels of the system~\cite{landauer,buttiker}. Clearly, the whole set $\tn$ cannot be extracted from the total
    conductance which only yields information on $\sum_n \tau_n$.
    The possibility of an accurate determination of the individual $\tn$
    was demonstrated by Scheer \textit{et al.}~\cite{Scheer98}
    for the  case of one-atom contacts, by analyzing electron transport
    in the superconducting state~\cite{Naveh}.
    This technique allowed to prove that the number of channels
    for one-atom contacts is basically determined by the valence orbital
    structure while the particular values in the set $\tn$
    depend also on the contact geometry at the atomic scale~\cite{review}.

    A natural question that arises is how the set
    $\tn$ evolves as the size of the contact is increased. One
    would expect that for sufficiently large contacts (where the
    size is larger than the mean free path) the transmission
    coefficients would be distributed according to
    $P(T)= \langle \sum_n \delta \left(T - \tau_n \right)
    \rangle = \langle G \rangle/ \left( 2G_0 T \sqrt{1-T}\right)$,
    \textit{i.e.} the universal
    distribution function predicted for a quantum conductor in
    the diffusive regime~\cite{dorokhov82,Beenakker97}.
    Although indirect evidence of this distribution has been
    obtained through shot-noise measurements~\cite{shotnoise}
    a direct determination of $P(T)$ in this regime is still an
    open experimental challenge.

    In the present work we combine experimental and theoretical
    efforts in  order to analyze  the evolution of the channel
    distribution in nanocontacts as the size of the contact is
    increased from the atomic-size limit.
    We obtain the set $\tn $
    from transport measurements in lead nanocontacts with
    conductances ranging from 1 to 15 $G_0$.
    These results are compared with model calculations in which the effects of
    geometry, atomic disorder, and band structure can be included.
    We find that even for small contacts ($G \sim $3--4 $G_0$) the
    channels distribution for Pb is
    unexpectedly close to the diffusive limit. We show that this
    behavior can be attributed to the particular band structure of
    Pb.

    Highly stable atomic scale contacts are formed  using a
    mechanically controlled break-junction (MCBJ)~\cite{muller}. A notched lead wire 99.99\% pure is glued on top of
    a flexible
    substrate. The rupture of the wire is carried out at 1.6 K in He exchange
    gas, thus preventing the freshly exposed surfaces from oxidizing.
    The contact can be
    reestablished and broken with picometer
    displacement resolution and stability.

    %
    %

    It was shown that in one-atom contacts the $\tn$ can be obtained
    from the experimental current-voltage characteristic curve
    (\textit{IV}) by fitting to the sum of $N$ one-channel \iv curves,
    \mbox{  $I(V)= \sum_{n=1}^N i(\tau_n,V)$}~\cite{Scheer97,Scheer98},
    which have been calculated elsewhere for arbitrary transmission $\tau$~\cite{subgap}.


    Scheer \textit{et al.}~\cite{Scheer97,Scheer98} determined the
    transmission of a small number of channels for one-atom
    contacts using a steepest descent based method~\cite{privatecomm.}.
    For larger contacts we have found that a simulated
    annealing algorithm is much more efficient.
    It is faster, a fact that becomes increasingly important
    as the number of channels to fit grows, and has the advantage
    of avoiding getting trapped at local minima.
    We use as acceptance criterion the root mean
    square deviation from the measured \iv curve, $\chi^2$.
    The fit starts from a random initial position in the
    $N$-dimensional space spanned by $\tn$. Then a random
    walk is performed. At each step the new fit is accepted
    if $\chi^2$ is smaller than the previous one.
    If $\chi^2$ is larger the step can
    still be accepted, according to a Boltzmann-type factor,
    with a pseudo-temperature which is lowered in a smooth way.
    $N$ is chosen large enough for the fit to result in several closed channels ($\tau <0.01$).

    %
    %
    \begin{figure}
    \includegraphics[width=\columnwidth]{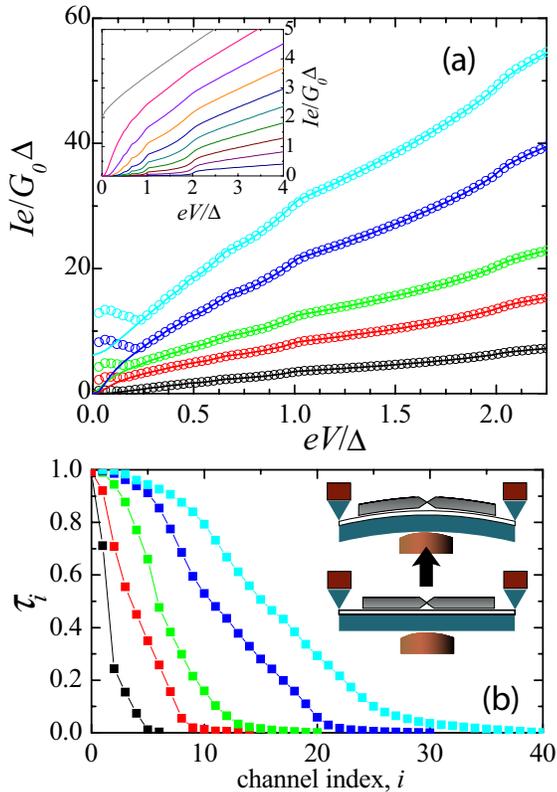}
            \caption{(Color on-line). Top panel shows experimental $\iv$ curves ($\circ$)
                in the superconducting state, for
                contacts of  $G = 2,4,6,10,15 \, G_0$, and the
                fits (lines).
                The value of the superconducting gap $\Delta=1.35$
                meV used in the fit is obtained from an \iv curve in the tunneling regime.
                The inset in a) shows the theoretical $\iv$ curves for a single channel
                used in the fitting procedure (taken from~\cite{subgap}).
                Bottom panel: %
                corresponding sets $\tn$ obtained from  the fits.
                The inset in b) shows a schematic representation of the
                MCBJ technique.
                }
      \label{expchannels}
    \end{figure}
    %

    We show in fig. 1 a) \iv curves for different contact realizations.
    These \iv curves exhibit subharmonic gap  structure for voltages below $2
    \Delta /e$ due to multiple Andreev reflections~\cite{subgap}.
    At low voltages the \iv curves show a
    contribution from the supercurrent peak and is not taken into
    account.

    In fig. 1 b) we show the channel transmissions obtained from the
    fit to the \iv curves in fig. 1 a). Channel indexes are ordered in decreasing transmission.
    An estimate of $P_{\mathrm{norm}}(T)=P(T) G_0/\langle G \rangle$ can be obtained
    from a finite number of experimental realizations in the following way.
    A histogram of the $\tn$ values is constructed dividing the transmission axis into
    14 equally sized bins. The $\tau$ values falling within the range of each bin are collected
    from 20 contact realizations of a given conductance.
    The result is normalized dividing by the conductance times the number of realizations.
    We find that $P_{\mathrm{norm}}(T)$ is bimodal, with a significant number of channels with
    high transmission.

    %
    %
    \begin{figure}
      \includegraphics[width=\columnwidth]{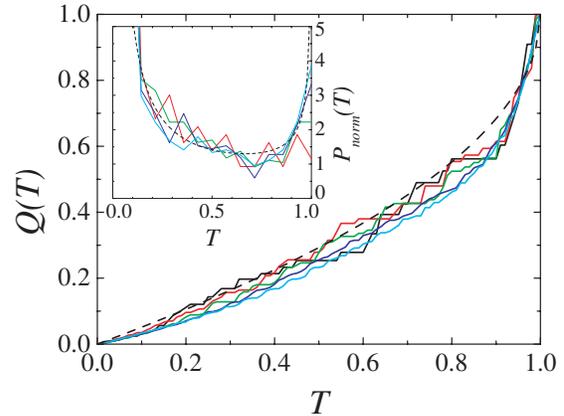}
      \caption{%
                (Color on-line). Estimate of $Q(T)$ for contacts of  $G = 2,4,6,10,15 \, G_0$
                (black, red, green, blue and cyan respectively).
                The $Q(T)$ corresponding to the diffusive case is shown as a dashed line.
                Inset: estimated $P_{\mathrm{norm}}(T)$.%
                }
      \label{Qfig}
    \end{figure}

    The channel distribution can be also conveniently presented
    using the well-behaved integrated quantity~\cite{schep-bauer}:
        \begin{equation}
          \label{eqQ}
          Q(T)= {\frac{G_0}{\langle G \rangle}} \int _0^{T}
          \mathrm{d} 
          T' \,    P(T') T' \, ,
        \end{equation}
    which is a regular function of $T$. $Q(T)$ is
    the relative contribution to the conductance of all channels with $\tau <T$.

    We obtain an estimate of $Q(T)$ from 20 experimental
    realizations making a histogram with a bin size equal to 0.01.
    The $\tau$ values that are lower than $T$ from all realizations of a given
    conductance are collected and added. Then the result is normalized
    dividing by $\sum \tau_n$.
    For a perfect ballistic conductor this  function
    is zero except for $Q(1)=1$.
    The distribution shown in fig.~\ref{Qfig} is not very sensitive
    to the size of the contact and remarkably similar to the universal
    diffusive limit. This similarity is rather unexpected due to the small
    size of the contact with less than 20 atoms in cross section.


    Previous experience
    for the case of one-atom contacts has demonstrated that band
    structure effects play a major role in the determination of the set
    $\{\tau_n\}$ for a given element~\cite{us98,Scheer98,byge1,mingo,byge2}.
    It is thus interesting to
    analyze whether this is still the case for contacts in the
    intermediate range (larger than one atom but smaller than the mean free
    path). For this purpose we have performed model calculations
    of the set $\tn$ for idealized geometries of the
    nanocontacts.

\begin{figure}
    \includegraphics[width=\columnwidth]{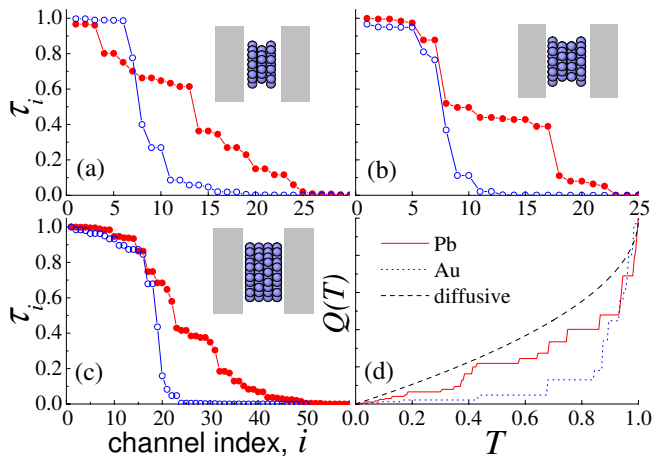}
    \caption{%
              (Color on-line). Distribution of transmission coefficients calculated
              for idealized geometries in which the number of
              atoms in the neck region layers are: a) 18-12-18,
              b) 18-12-12-18 and c) 31-27-27-31. In all figures the full and open
              circles correspond to Pb and Au respectively. In panel d) we show the
              distribution $Q(T)$ corresponding to the case of panel c).%
              }
    \label{theoretical-taus}
    \end{figure}

    As in previous studies we use a self-consistent parametrized
    tight-binding (TB) model in which the main features of the bulk
    band-structure are included~\cite{us98}.
    We shall consider model geometries
    like the ones depicted in the insets of fig.~\ref{theoretical-taus}
    in which the neck of the contact is represented by several atomic layers
    of different cross section grown along the (111) direction on a fcc
    lattice. This central region is connected to the left and right
    electrodes represented by perfect semi-infinite crystals.
    The corresponding TB Hamiltonian can be written as
    $\hat{H} = \sum h_{i,j,\alpha,\beta}
    \hat{c}^{\dagger}_{i,\alpha,\sigma} \hat{c}_{j,\beta,\sigma}$ $=$
    $\hat{H}_{c}$ $+$ $\hat{V}_L$ + $\hat{V}_R$ + $\hat{H}_L$ +
    $\hat{H}_R$, where $\hat{H}_{L,R}$ and $\hat{H}_{c}$ describe the
    uncoupled left and right electrodes and the central part of the contact
    respectively;
    $\hat{V}_{L,R}$ being a coupling term between the central region and the
    electrodes.
    The matrix elements $h_{i,j,\alpha,\beta}$, where $i,j$ and
    $\alpha,\beta$ design sites and orbitals respectively, are taken from
    fits to the bulk {\it ab-initio} band structure.  As a
    self-consistency condition we impose local charge neutrality
    on each site. Disorder in the atomic positions at the central region can
    be included by rescaling the hopping elements according to the distortion
    of the different bond length with respect to the bulk values%
    ~\cite{comment-hopping}.

    Once the TB Hamiltonian has been built the conductance $G(E)$ = $G_0
    \mbox{Tr} \left[\hat{t}^{\dagger}(E) \hat{t}(E) \right]$
    is calculated in terms of the matrix elements of the Green function
    operator $\hat{G}^{r}(E)$ =
    $\lim_{\eta\rightarrow 0} \left[E + i\eta - \hat{H}\right]^{-1}$ using
    :
    \begin{eqnarray}
    \hat{t}(E) = 2 \hat{\Gamma}^{1/2}_L(E) \hat{G}^{r}_{1N}(E)
    \hat{\Gamma}^{1/2}_R(E),
    \end{eqnarray}
    where $\hat{\Gamma}_{L,R}$ are the matrix tunneling rates connecting
    the central region to the leads~\cite{us98}.  The transmission matrix
    $\hat{t}^{\dagger}\hat{t}$ can then be diagonalized in order to obtain
    the transmission eigenchannels and eigenvalues $\tau_n$.

    %
    %


    In fig.~\ref{theoretical-taus} we show typical results for the
    transmission sets that are obtained for Pb contacts of intermediate
    cross section. The effect of an
    specific contact geometry is illustrated in the upper panels
    (fig.~\ref{theoretical-taus} a) and b)). As can be observed, a
    characteristic feature of Pb is the appearance of steps
    corresponding to partially open channels with similar
    transmission. For comparison we also show in this figure the
    corresponding results for Au contacts in order to remark the
    differences with a monovalent metal.
    Although the detailed structure of these curves is
    strongly dependent on the specific geometry, the behavior for Pb
    is clearly in contrast to the one observed in Au where
    partially open channels are rarely observed. These differences can
    be directly related to the particular band structure of both
    metals. While in the case of Au the channels can be
    associated with a single band of s-character at the Fermi
    energy, in Pb the channels arise from the contribution of several bands
    with sp-character. In fact, for the same geometry the total number
    of channels with significant transmission is larger for Pb than
    for Au. When the cross section of the contact is increased (fig.%
    ~\ref{theoretical-taus} c)) the partially open channels in Pb tend
    to define a rather continuous distribution which fairly resembles
    that of a diffusive conductor. In panel d) we compare the
    corresponding distributions $Q(T)$ with the result expected for
    a diffusive wire.

    As a result of these previous calculations one can conclude that,
    even in the absence of atomic disorder, Pb contacts deviate
    strongly from the ballistic behavior observed in the case of Au.

    %
    %
    \begin{figure}
    \includegraphics[width=\columnwidth]{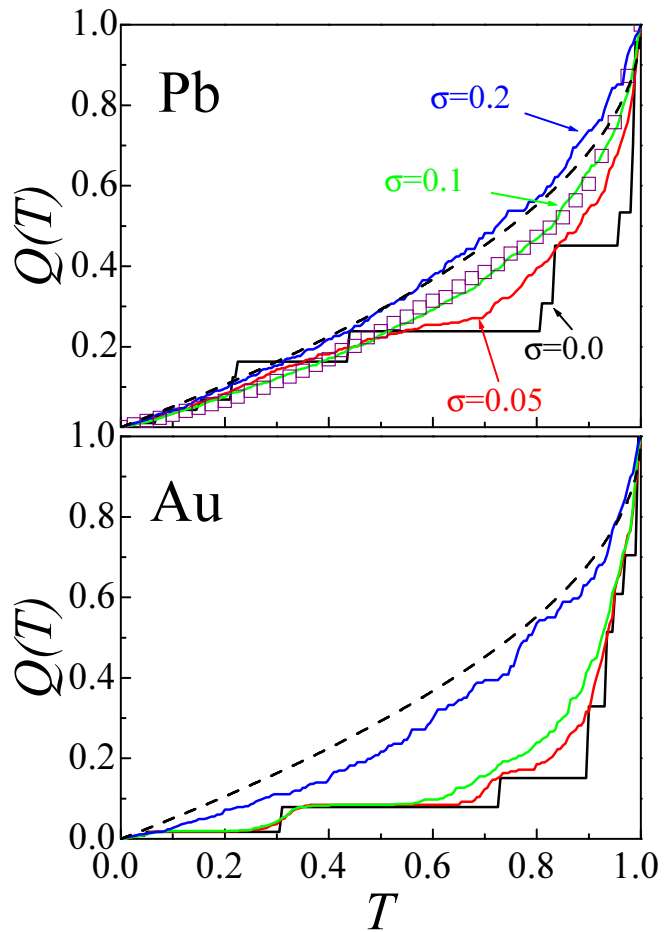}
    \caption{%
             (Color on-line). Effect of atomic disorder in the calculated distributions
             $Q(T)$ for Pb and Au. The curves correspond to different values of
             the disorder parameter $\sigma$ described in text: 0 (black line),
             0.05 (red), 0.1 (green) and 0.2 (blue). The symbols correspond to
             an average of the experimental results from $10$ to $15 G_0$. The
             dashed line corresponds to the universal diffusive distribution.}
    \label{theoretical-disorder}
    \end{figure}

    In a second step we have analyzed the influence of atomic disorder
    in the calculated distributions. Disorder is assumed to be localized
    in the neck region (we will discuss later the effect of disorder on
    the bulk electrodes). The degree of disorder is characterized by
    the mean square deviation of the bond lengths with respect to the
    idealized structure, $\sigma$. fig.~\ref{theoretical-disorder} shows the
    evolution of the distribution function $Q(T)$ as $\sigma$ is
    increased for a particular contact geometry.
    As can be observed for Pb the distribution tends rather quickly
    to the diffusive one with increasing disorder.
    We also show  the experimental distribution in which several cases with
    total conductances ranging from 10 to 15 $G_0$ have been averaged.
    This distribution is in fairly good agreement
    with the calculated one for the case $\sigma=0.1$, which roughly
    corresponds to a mean free path $l_\mathrm{e} \sim 10$ nm.
    This rather low degree of disorder is consistent with what is known  both from
molecular-dynamics simulations and images made by transmission
electron microscopy of metallic nanocontacts~\cite{review}.

    It should be stressed that these distributions
    still exhibit some deviations with respect to the universal
    curve, specially for high transmission. We can associate these
    deviations with the fact that the size of the contact is still
    considerably smaller than the mean free path.

    In contrast to the case of Pb, Au exhibits a clear deviation from the
    universal distribution even for the larger values of $\sigma$
    considered. Notice that $\sigma = 0.2$ already corresponds to
    an unexpectedly high degree of disorder~\cite{comment-disorder}.

    As a final issue we estimate the effect of disorder on the bulk
    electrodes by means of a phenomenological model. In this model
    the electrodes are diffusive conductors characterized by a classical
    probability $P_{\mathrm{cl}}(t)$ for returning to the contact region after a
    given time $t$.
    Modelling the electrode as a cone with openning angle $\gamma$ this
    probabilitiy is given by $P_{\mathrm{cl}}(t) =     v_\mathrm{F}/\left\{2\sqrt{3\pi}k^2_\mathrm{F}\left(D
    t\right)^{3/2} \left(1 - \cos{\gamma}\right)\right\}$,
    where $k_F$ and $v_F$ are the Fermi wavevector and velocity%
    respectively and $D = v_\mathrm{F} l_\mathrm{e}/3$ is the diffusive constant
    ~\cite{ludoph99}. Following the arguments of ref.~\cite{ludoph99}
    the correction to the transmission distribution can be written as

    \begin{equation}
    \delta P(T)= \langle \sum_n \delta \left( T -
    \tau_n \right )\left(1 - \tau_n \right) \rangle
    p_{\mathrm{return}} \, ,
    \end{equation}

    where $\tau_\mathrm{e} = l_\mathrm{e}/v_\mathrm{F}$ is the elastic scattering time
    and $p_{\mathrm{return}} =  \int _{\tau_{\mathrm{e}}} ^{\infty} P_{\mathrm{cl}}(t) \mathrm{d} t$.
    Assuming that typically $\gamma \sim 45^{\circ}$ and that $l_\mathrm{e}$ can range from
    $\sim 10$ nm to $\sim 4$ nm in Pb, one obtains $p_{\mathrm{return}} \sim
    0.005 - 0.02$ leading to a rather small correction
    which hardly modifies the theoretical results
    already shown in fig.~\ref{theoretical-disorder}.

    In conclusion, we have presented a combined experimental and theoretical
    analysis of the conduction channel distribution of nanoscale contacts
    with several atoms in cross section. The experimental method demonstrates
    the applicability of the technique developed by Scheer \textit{et al.}%
    ~\cite{Scheer97,Scheer98} to much larger contacts. It is found that
    for Pb contacts the distribution tends rather fast to the universal
    diffusive limit. We show that this behavior can be associated with
    the specific band structure of Pb which favors the appearance of
    partially open channels in contrast to what is predicted for monovalent
    metals like Au.

    \acknowledgements
    We thank S. Vieira for useful comments and continuous support. This work has been financed       by  the  MCyT  under project
          MAT2002-11982-E within the EUROCORES Programme SONS of the European
          Science Foundation, which is also supported by the European
          Commission, Sixth Framework Programme and by  MCyT (BFM2001 0150 and
         MAT2001 1281), GR/MAT/0111/2004 and CAM (07N/0053/2002 and
         07N/0039/2002).

\end{document}